# Effect of ice nucleating proteins on the structure-property relationships of ice: A molecular dynamics study

Ali K. Shargh[1*], Christopher D. Stiles[1,2], Jaafar A. El-Awady[1*]

[1] Department of Mechanical Engineering, Johns Hopkins University, Baltimore, Maryland 21218, U.S.A.

[2] Research and Exploratory Development Department, Johns Hopkins Applied Physics Laboratory, Laurel, Maryland 20723, U.S.A.

# Abstract

Ice-nucleating proteins (INPs) are a unique class of biological macromolecules that catalyze the freezing of supercooled water far more efficiently than homogeneous nucleation. Their remarkable efficiency has motivated applications across diverse sectors, including agricultural frost protection, food processing and packaging, biomedical cryopreservation, and even strategies for mitigating glacier ice loss. The ice-nucleation performance of INPs and the mechanical behavior of the ice they produce depend strongly on their structural and biochemical characteristics. However, the links between INP properties, the resulting ice microstructure, and their mechanical behavior have yet to be systematically established. In this study, coarse-grained molecular dynamics (CGMD) simulations using the machine-learned ML-BOP potential are employed to investigate how varying INP densities influence the ice nucleation temperature, the resulting ice microstructure,

---

[*] Corresponding authors:
 Email addresses: ashargh1@jhu.edu (A. K. Shargh), jelawady@jhu.edu (J. A. El-Awady)

and the mechanical behavior of the formed ice under creep tensile loading. We find that, depending on their density, INPs can significantly raise the ice nucleation rate while altering the grain structure of ice. Our simulations reveal that INP-assisted nucleation leads to faster stabilization of the resulting polycrystalline ice composed of hexagonal ice (ice Ih) and cubic ice (ice Ic) as compared to nucleation in pure water. Moreover, higher INP densities and smaller ice grain sizes reduce the overall yield stress, while promoting diffusion-accommodated grain boundary sliding creep. These findings provide molecular-level insight into how INPs influence both the nucleation process and the mechanical behavior of ice, highlighting a pathway to engineer ice with tailored stability for real-world settings, including human activities and infrastructure in polar and icy environments.

# Keywords

Molecular dynamics simulations, Ice-nucleating proteins, Creep behavior, Ice nucleation, Polycrystalline ice

# 1. Introduction

Ice nucleating proteins (INPs) are a remarkable class of biological macromolecules that catalyze the freezing of supercooled water at rates far higher than would occur under homogeneous nucleation[1]. Found in certain bacteria such as Pseudomonas syringae, they have evolved specialized ice-binding sites capable of templating the arrangement of water molecules into ice-like structures[2,3]. Because of their ability to control ice formation with high efficiency, INPs have been explored in diverse fields, including agricultural frost protection, controlled freezing in food processing and packaging, and biomedical cryopreservation[4,5]. In addition, INPs enable freeze-tolerant organisms such as winter rye, citrus fruit, and insects to survive at low supercooling temperatures by initiating ice formation at relatively high subzero temperatures, thereby reducing the risk of cellular damage caused by rapid or uncontrolled freezing[6].

Beyond these applications, INPs have also been proposed as part of emerging strategies to mitigate the ongoing loss of polar ice sheets[7]. The polar ice sheets hold nearly two-thirds of Earth's freshwater ice[8], and changes in their mass can drive major shifts in sea level and ocean circulation. Satellite observations over recent decades confirm accelerating mass loss from both the Greenland and Antarctic ice sheets, contributing significantly to global sea level rise[9]. While reducing greenhouse gas emissions remains one of the most direct way to slow ice-sheet loss[10], there is growing interest in complementary interventions. These range from large-scale engineering solutions, such as installing underwater artificial sills to block warm ocean currents[11,12], to surface-based strategies, such as deploying reflective materials and automated seawater pumping to enhance local ice thickening[13,14]. In this context, INPs offer a fundamentally different route, acting at the molecular scale to alter the freezing behavior of surface meltwater and potentially influence ice-sheet dynamics.

While engineered INPs offer exciting potential, their performance depends heavily on a range of structural and biochemical factors such as density of INPs, orientation, sequence-specific motifs, and local interactions with $H_2O$ molecules. Understanding how these microstructural parameters influence ice nucleation efficiency is essential for the rational design of robust, high-performance

INPs tailored for targeted applications such as frost protection, cryopreservation, or mitigating the ongoing loss of polar ice sheets.

However, establishing these relationships experimentally is challenging due to the nanoscale nature of the nucleation process, the complexity of protein–water interactions, and the considerable effort required to test across multiple conditions. Molecular dynamics (MD) simulations offer a robust and broadly applicable framework to investigate the connections between microscopic structure and macroscopic behavior across a wide range of material systems, providing insights that can inform and refine subsequent experimental studies[15–17]. Consequently, several studies have employed MD simulations to investigate how INPs facilitate ice nucleation at the atomic scale[18–26]. For example, Roeters et al.[22] unveiled that INPs adopt a β-helical structure at water interfaces, where their interaction with surrounding water promotes structural ordering in the interfacial hydrogen-bond network of liquid water. The authors found that this ordering intensifies as temperature approaches the melting point, enhancing ice-nucleation efficiency. In a different study, Hudait et al[18] revealed that INPs recognize and bind to ice through diverse interfacial motifs, including anchored clathrate and ice-like structures. Their findings highlighted that multiple structural pathways could support effective ice binding, beyond a single binding configuration. Most recently, Alsante et al.[26] demonstrated that ice nucleation efficiency of the proteins was independent of molecular weight, and that higher concentration or aggregation did not consistently lead to enhanced nucleation activity.

Despite these important advances in our understanding, questions such as the influence of INP spatial density and distribution on ice nucleation remain unanswered. Furthermore, earlier studies have been conducted from a chemistry-focused perspective, leaving a gap in systematically understanding how variations in INP microstructure affect the morphology and grain structure of the resulting ice, which ultimately control their structural and mechanical properties. Importantly, ice generated through INP activity is not just a thermodynamic phase, but a material with mechanical properties, where its performance in terms of strength, and deformability can determine the stability of ice in natural and engineered environments. Understanding these mechanical aspects is critical for the durability and resilience of ice in real-world settings, including ice-based infrastructure, polar research stations and surface operations, where ice stability is essential.

As such, given that ice ultimately deforms and fails under thermo-mechanical stress, such as creep observed in glaciers or cracking during cryogenic preservation[27,28], understanding how INPs influence the mechanical behavior of the ice they nucleate is of great importance. Interestingly, MD simulations have been used extensively to investigate the mechanical behavior of ice[29–36]. Nevertheless, these studies have solely focused on pure ice, and the influence of INPs on the mechanical behavior of ice remains largely unexplored.

In this study, we employ MD simulations to investigate how the density of ice-nucleating proteins (INPs) influences their ice nucleation efficiency and the resulting ice microstructure. We then examine how variations in INP density affect the mechanical behavior of the formed ice under creep loading. By addressing both nucleation efficiency and the resulting mechanical properties, this study aims to link molecular-scale interactions to macroscale ice stability, providing insights that could guide the design of ice with tailored durability and resilience in real-world settings, including polar operations, engineered ice structures, and other environments where ice stability is critical.

The remainder of the paper is organized as follows. Section 2 describes the numerical methodology, including the generation of ice matrices with varying INP densities and the MD simulation setup. The results and discussions are presented in Section 3, where Section 3.1 examines the effect of INP density on the ice nucleation temperature. Section 3.2 compares microstructural evolution during homogeneous and heterogeneous nucleation. Section 3.3 discusses the mechanical response of ice as a function of INP density under creep loading. Finally, Section 4 summarizes the key findings and conclusions of this work.

## 2. Methods

### 2.1. Simulation setup

Molecular dynamics simulations are performed using the open-source Large-scale Atomic/Molecular Massively Parallel Simulator (LAMMPS)[37]. Visualization and analysis of the simulation results are carried out using OVITO[38]. Periodic boundary conditions are applied along all three directions. In our simulations, water molecules are modeled at the coarse-grained level, with each water molecule represented by a single bead. As a model INPs, we focus on the Pseudomonas syringae (PsINP), which serves as a representative system for studying ice-binding behavior[25]. PsINP repeats a 16-amino-acid sequence (GYGSTQTSGSESSLTA) derived from the central ice-binding domain of the 1INAZ.pdb structure, which we use without modification[39]. We model PsINP using a united-atom representation, as shown in Figure 1, in which all non-hydrogen atoms are explicitly included. The 3-body ML_BOP[40] potential is chosen from the available interatomic potentials[40–42] for modeling water-water interaction. The ability of this potential to describe the structure and thermodynamics of both water and ice, as well as the mechanical behavior of ice under loading, has been previously validated[35,42–45]. For modeling protein-water interactions, we use the Stillinger–Weber potential[46] with parameters provided by Hudait et al[18]. The software Chimera[47] is used to differentiate between particle types such as carbonyl, amide, and amine groups, which are necessary for specifying the interatomic potential. The protein is treated as a rigid body, with no internal degrees of freedom, which means that the forces between pairs of atoms in the protein are ignored. This assumption is necessary to preserve the protein's structural stability, since, to our knowledge, no potential in the united-atom model is yet available that accurately captures the internal or inter-protein interactions.

The equations of motion are integrated using the velocity Verlet algorithm with a time step of 5 fs. Temperature and pressure are controlled via the Nosé–Hoover thermostat and barostat, with damping constants of 0.5 ps and 2.5 ps, respectively. Hexagonal (Ih), cubic (Ic), and amorphous/liquid ice phases are identified using the CHILL+ structure identification algorithm[48], as implemented in OVITO. Individual grains and their size distribution are identified using the Graph Clustering algorithm implemented in OVITO, with 'Polyhedral Template Matching' enabled to determine lattice orientations and 'CHILL+' activated to characterize different crystalline structures, with all parameters set to their default values.

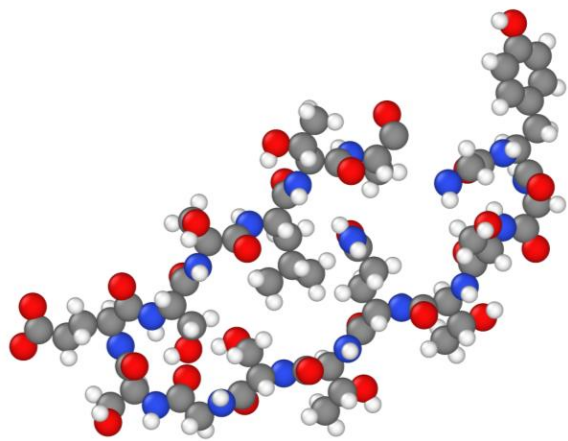

**Figure 1:** All-Atom (AA) representation of one-loop PsINP. Carbon atoms are shown in gray, nitrogen in blue, oxygen in red, and hydrogen in white. In the current MD simulations, the INPs are modeled at the united-atom level, where hydrogen atoms are not considered.

## 2.2. Microstructure generation

To study how the density of INPs influences ice nucleation and the ice's subsequent mechanical response, we first create a simulation box of size 26 × 26 × 28 nm³ containing 511,980 randomly distributed water molecules to represent the liquid phase. We then place INPs in the simulation box using two different configurations. In the first configuration, a single INP, either 2.75, 5.5, or 11 nm in length, is placed at the center of the simulation box with its central axis aligned along the x-axis. Different lengths of INPs are obtained by replicating the PsINP segment (extracted from the 1INAZ.pdb structure described earlier) along its central axis. In the second configuration, 2, 5, 10, and 15 INPs, each 5.5 nm long, are randomly distributed in the simulation box using a custom Python script. In this random sampling approach, the position of each INP is generated using Poisson disk sampling, with a minimum separation of 1nm enforced between any pair of atoms from two different INPs to prevent overlap. Additionally, the orientation of each INP is randomly rotated in space using quaternion-based uniform sampling to capture orientation variability. In the remainder of the paper, a simulation with a single 5.5 nm INP corresponds to INP density = 1 L, a simulation with a single 2.75 nm INP corresponds to INP density = 0.5 L, and a simulation with a single 11 nm INP corresponds to INP density = 2 L. The polycrystalline ice Ih simulations discussed in Section 3.3 also have a simulation cell size of 26 × 26 × 28 nm³ and contain 32, 219, or 478 grains. These atomic structures are generated using Atomsk, which is an open-source command-line program for atomic structure generation[49].

## 2.3. Dynamic cooling and loading protocols

To obtain the ice nucleation temperature for each simulation setup, the atomic structure is first minimized to a local potential energy minimum using the conjugate gradient algorithm with a maximum force tolerance of $10^{-6}$ eV/Angstrom. The system is then cooled down in a constant number of particles, constant pressure, and constant temperature (NPT) ensemble[50] using a cooling rate of 1 K.ns$^{-1}$. While this cooling rate is significantly higher than typical experimental conditions (e.g., microliter droplets are cooled at approximately 1 K.min$^{-1}$)[51], this cooling rate is within the range commonly used in previous MD studies[18,40]. This high cooling rate is essential to observe nucleation events within the limited timescales accessible to MD simulations. Once the largest nucleus is observed to be of size ≥ 100 molecules[40], the cooling simulation is stopped and the corresponding temperature is recorded as the ice nucleation temperature. The supercooled liquid water containing the stable ice nucleus is then kept at the corresponding ice nucleation temperature for 130 ns in an NPT ensemble to obtain the final well-equilibrated microstructure. Tensile loading simulations are then conducted as follows. First, uniaxial tensile loading is performed in an NPT ensemble using a strain rate of $5 \times 10^8$ s$^{-1}$ to determine the yield stress, $\sigma_y$. Importantly, although the strain rate of $5 \times 10^8$ s$^{-1}$ is higher than typical experimental rates, it is in the standard range of strain rates in MD simulations. This is due to MD simulations being limited by the timescales they can reach. It is important to note that these MD simulations are not intended to replicate experimental timescales but rather to provide insight into deformation mechanisms and relative material behavior. Tensile creep simulations are then conducted by applying a constant uniaxial stress ranging from $0.4\sigma_y$ to $0.8\sigma_y$ along the x-direction, while maintaining zero stress in the lateral directions, all within an NPT ensemble[52].

# 3. Results and Discussion

## 3.1. Effect of INP density on the ice nucleation temperature

To investigate the effect of INP density on the ice nucleation temperature, we construct several models containing liquid water with varying INP densities and perform cooling simulations following the steps described in Section 2.3. The variation of ice nucleation temperature with INP density is shown in Figure 2(a). The error bars represent the mean and standard deviation from three independent simulations performed with different random velocity seeds. For simulations containing multiple INPs, randomization also includes variations in the location and orientation of the INPs. For comparison, the ice nucleation temperature of pure liquid water without INP is also included. It is observed from Figure 2(a) that within the length and time scales considered in this study, the inclusion of INPs can improve the ice nucleation temperature of liquid water by up to 23K from a homogeneous baseline of $206 \pm 1.7$ K. This enhancement aligns well with the 19.6 K increase observed experimentally for Snomax, which contains the InaZ protein[53]. Notably, direct calculation of nucleation rates from MD simulations requires specialized rare-event sampling methods[54] and is beyond the scope of this study. However, the observed 23 K enhancement in nucleation temperature relative to homogeneous nucleation implies a substantial increase in the nucleation rate. According to the classical nucleation theory, nucleation on INP surfaces introduces a geometric factor $f(\theta) < 1$ that lowers the free energy barrier for critical nucleus formation, thereby accelerating nucleation kinetics. Because the nucleation rate depends exponentially on this barrier, heterogeneous nucleation results in a substantial increase in the nucleation rate compared with homogeneous nucleation under comparable cooling conditions.

It is important that the freezing of water near 273 K during cooling cannot be captured in MD simulations due to two primary limitations: (1) the high cooling rates used in MD simulations suppress nucleation relative to slower experimental rates, and (2) the small MD simulation box sizes reduce the probability of forming a critical nucleus. Despite these limitations, MD simulations give a fundamental understanding of the nucleation mechanisms and effect of INPs on ice nucleation.

As shown in Figure 2(a), the nucleation temperature initially increases, reaches a plateau around 229K - 230K, and then decreases with further increase in INP density. The plateau can be attributed to the fact that the finite width of the INP's binding site limits the lateral size of the ice nucleus that it can effectively stabilize[55]. As a result, further increasing the length of INP does not significantly reduce the nucleation barrier. At higher INP densities shown in Figure 2(a), the reduced spacing between INPs cause their ice-nucleating regions to overlap, disrupt the surrounding water structure, and reduce the free volume between them needed to stabilize a critical ice embryo. These effects collectively reduce nucleation efficiency, resulting in a lower nucleation temperature. We note that similar nucleation temperatures are observed for a single INP of length 2L and two INPs of length L in Figure 2(a), which suggests that increasing INP density either by extending their length or increasing their number yields a comparable ice nucleation temperature response.

We also observe that all ice microstructures formed by cooling liquid water with varying INP densities exhibit a polycrystalline structure composed of both hexagonal ice Ih and cubic ice Ic. To investigate how ice microstructures are influenced by the initial INP density and the resulting effects on mechanical performance, we analyze additional features of the simulation results, including the Ih/Ic phase fraction, average grain size, and the number of grains in the formed ice after the equilibration step, as summarized in Figure 2(b)–2(d).

As observed in Figure 2(b), the Ih/Ic ratio generally decreases with increasing INP density, with the highest Ih content observed in the ice formed without INP. This suggests that INPs preferentially lower the nucleation barrier for ice Ic. Previous MD studies have reported that certain surfaces can promote ice Ic formation by matching its structural motifs[56], implying that INP surfaces may similarly favor Ic nucleation over Ih. Interestingly, the uncertainty associated with the simulation containing two INPs is significantly higher than that of the single-INP case of length L shown in Figure 2(b). This increased variability arises from the fact that, in the three independent trials of the two-INP configuration, the relative distance between the INPs are different. Since the relative spacing between the INPs can significantly influence the local nucleation environment and the resulting Ic/Ih fraction, these variations in INP separation result in a broader distribution of outcomes and higher statistical uncertainty.

Next, the evolution of the number of grains is shown in Figure 2(c). As INP density increases, a two-stage trend is observed. Initially, increasing the INP length leads to a reduction in the number of grains. However, with a further increase in INP density, the number of grains increases. At low to moderate INP densities, increasing the INP length provides larger continuous nucleation sites, promoting crystal growth from fewer nuclei and thus reducing the number of grains. However, as the INP density further increases through adding more numbers of INPs, the number of nucleation sites becomes sufficiently high for multiple grains to nucleate independently and simultaneously, resulting in a higher number of grains. As expected, the evolution of the average grain size shown in Figure 2(d) exhibits an inverse trend compared to the number of grains shown in Figure 2(c).

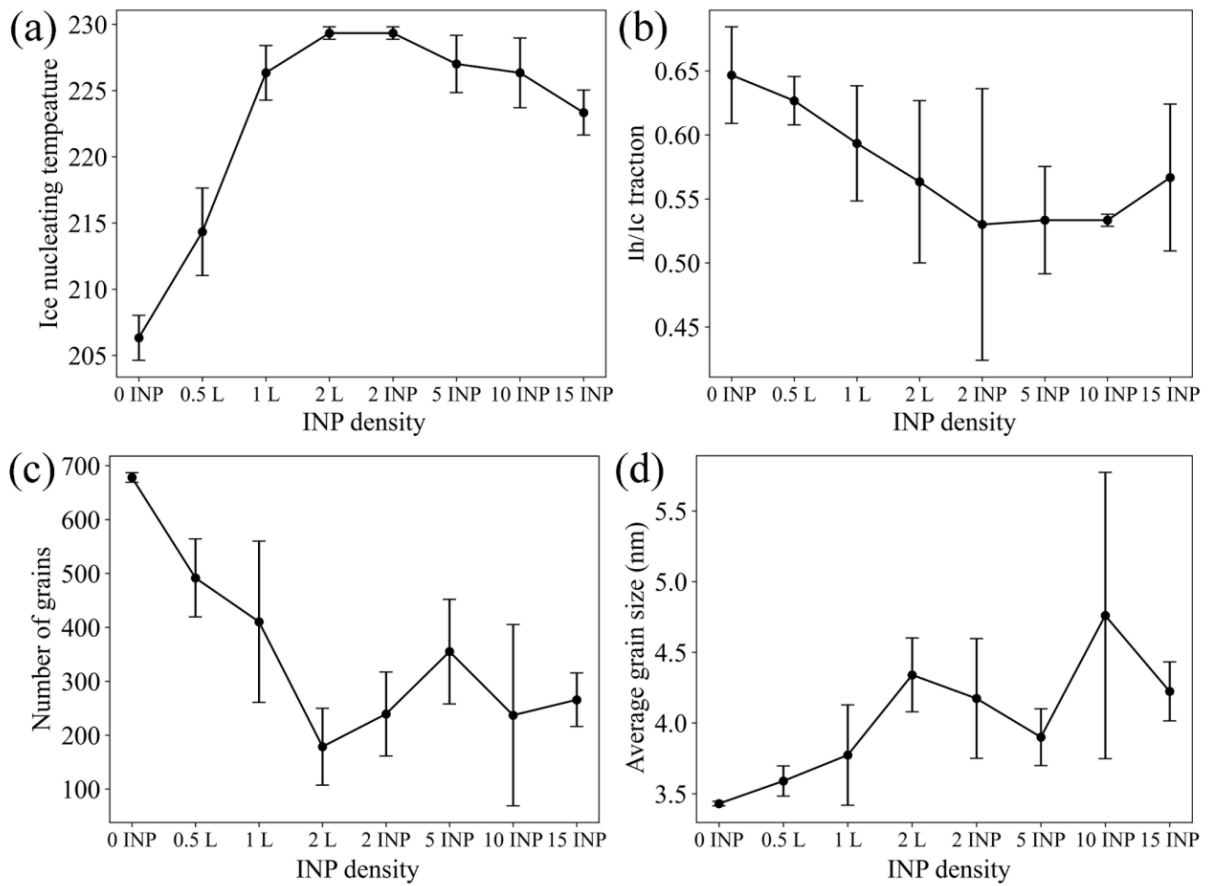

**Figure 2:** Correlation of density of INPs with ice nucleation temperature, Ih/Ic fraction, average grain size, and the number of grains in the final well-equilibrated structures investigated in this study following a 130 ns equilibration.

To further investigate the nanoscale microstructural evolution during equilibration, the following section extends our investigation by comparing how ice microstructure develops during the freezing of pure liquid water (homogeneous nucleation) and water containing a single INP with size L (heterogeneous nucleation).

## 3.2. Microstructural evolution during homogeneous versus heterogeneous nucleation

To compare microstructural evolution during homogeneous and heterogeneous nucleation, Figure 3 shows the process of ice nucleation and the subsequent ice formation in pure liquid water versus water containing a single INP. In this figure, the snapshot at $t = 0$ corresponds to the formation of stable ice nuclei. For clarity, only ice Ih particles (colored cyan) and ice Ic particles (colored orange) are shown. As shown in Figure 3(a), homogeneous ice nucleation leads to the formation of multiple nuclei dispersed throughout the liquid. As equilibration proceeds, these nuclei grow and merge, resulting in a polycrystalline ice structure. Consequently, cooling pure liquid water without INPs (i.e., homogeneous nucleation) results in an ice structure with a greater number of grains compared to ice formed in the presence of INPs (i.e., heterogeneous nucleation), as seen in Figure 2(c). In contrast, heterogeneous nucleation shown in Figure 3(b) begins with the formation of a single stable nucleus on the INP, which grows continuously through the surrounding liquid during equilibration. This results in a polycrystalline structure with a lower number of grains than that formed via homogeneous nucleation, as shown in Figure 2(c).

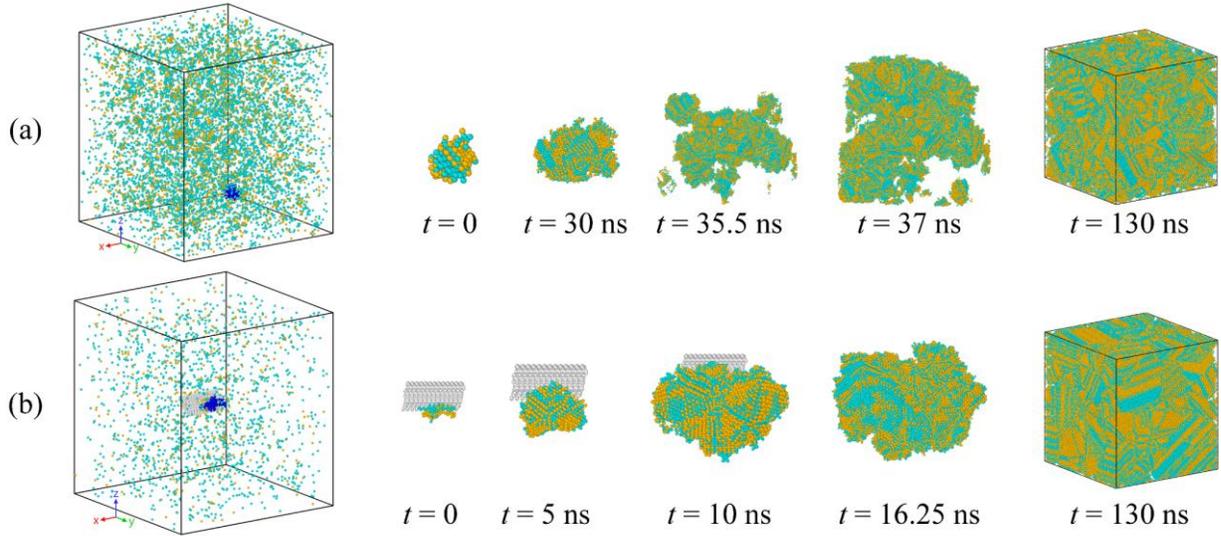

**Figure 3:** Process of ice nucleation and the subsequent ice formation in: (a) pure liquid water; and (b) liquid water containing one INP with a size of L under a cooling rate of 1 K/ns and constant temperature of 208 K and 226 K, respectively. The far-left image shows the simulation cell after the formation of a stable ice nucleus, highlighted in blue. This instance is designated as time $t = 0$. The zoom-in images show the temporal evolution of the cluster. For clarity, only the particles belonging to the INP (colored gray) and those ice particles in the ice Ih (colored cyan) and ice Ic (colored orange) states are shown. The number of $H_2O$ molecules in the ice phase at each time step are: (a) t = 0 ns (Ic: 2931, Ih: 6204), 30 ns (Ic: 43035, Ih: 32743), 35.5 ns (Ic: 67260, Ih: 49446), 37 ns (Ic: 75516, Ih: 54601), 130 ns (Ic: 205161, Ih: 127565); (b) t = 0 ns (Ic: 468, Ih: 1472), 5 ns (Ic: 1114, Ih: 1826), 10 ns (Ic: 7037, Ih: 4756), 16.25 ns (Ic: 36490, Ih: 23978), 130 ns (Ic: 247300, Ih: 152576).

To better highlight the differences in the evolution of ice microstructure formed heterogeneously from the INP surface compared to that formed via homogeneous nucleation, Figure 4 shows the time-dependent variation of the average grain size and number of grains during equilibration. In this figure, $t = 0$ corresponds to the first timestep in which 50% of the particles in the simulation cell are characterized crystalline by the Chil+ algorithm in OVITO, while the remainder are classified as 'other'-type by the algorithm.

A comparison between Figure 4(a) and 4(b) reveals that the number of grains stabilizes more quickly following heterogeneous nucleation than homogenous nucleation. This is because the lower nucleation barrier at the INP surface allows nucleation to begin earlier at higher temperatures, leading to the formation of a dominant stable nucleus that rapidly grows to fill the system and suppresses additional nucleation events, thereby accelerating the stabilization of the number of grains. Notably, while the number of grains in homogeneous nucleation shows minimal variation after $t = 50$ ns, heterogeneous nucleation exhibits a slight continuous decrease in the number of grains within the timescale investigated in this work, even though the corresponding grain size shows a minimal change. We have further extended the simulations to longer timescales (see Supplementary Figure S1), and this observed trend has persisted, suggesting that ongoing grain growth occur.

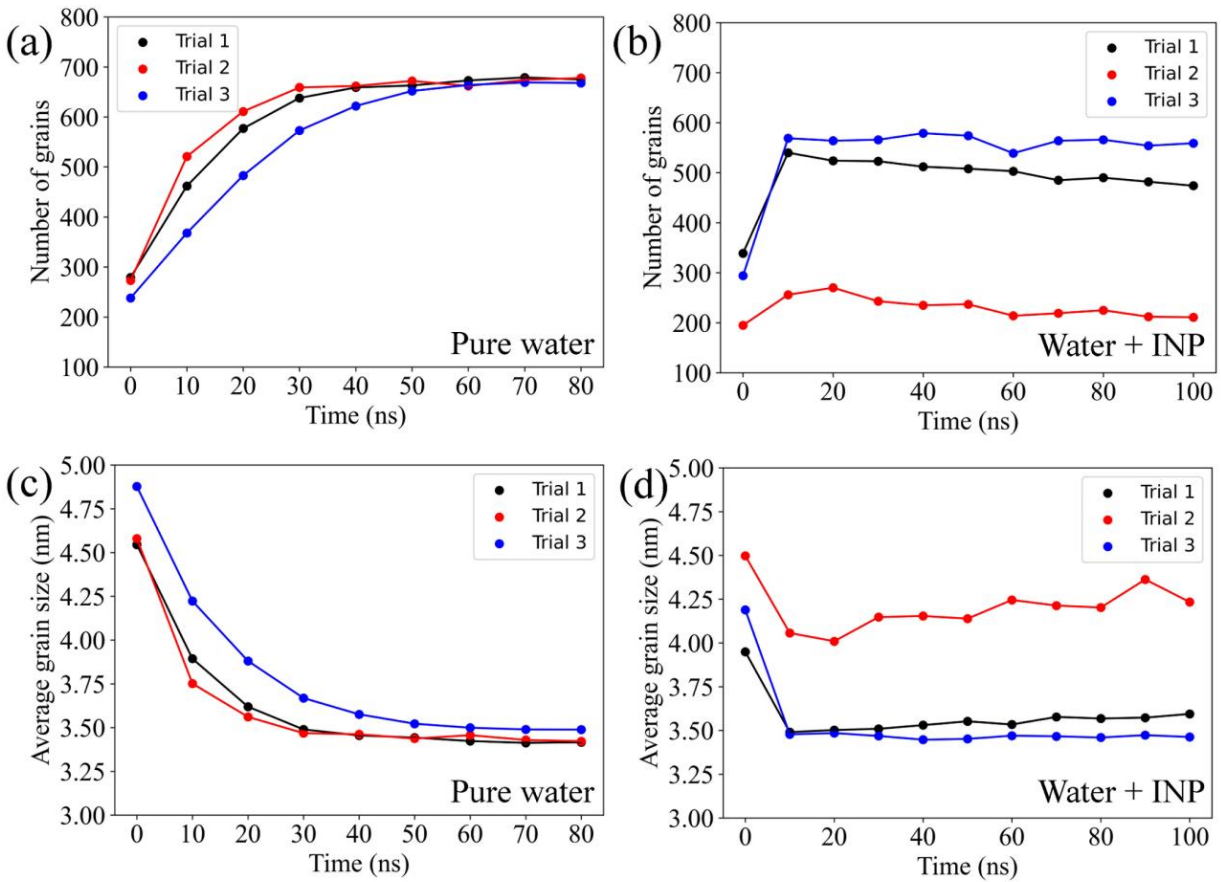

**Figure 4:** (a-b) shows the evolution of the number of grains in pure water and water with one INP having size L, respectively. (c-d) shows the evolution of the average grain size in pure water and water with one INP having size L, respectively. Time $t = 0$ represents the first simulation time at which 50% of the particles in the simulation cell are characterized crystalline with the ChiI+ algorithm in OVETO.

Given that ice Ih is thermodynamically more stable than ice Ic, one might hypothesize that an Ic-to-Ih phase transformation drives the gradual decrease in the number of grains. To further investigate this, we evaluate the evolution of the fraction of both ice Ih and Ic after homogeneous and heterogeneous nucleation. As shown in Figure 5, both Ih and Ic fractions remain mostly constant within the steady-state region over the timescale investigated, with only minimal fluctuations. This stability indicates that no significant phase transformation occurs within the timescale of our MD simulations, thereby ruling out Ic-to-Ih transformation as the cause of the observed decrease in the number of grains. Nevertheless, it should be noted that the experimentally observed thermodynamically stable phase of water below 0 °C at 1 atm is ice Ih[57]. Although our MD simulations cannot capture the Ic-to-Ih transformation due to the limitations of the accessible MD timescales, such transitions are expected to occur over much longer times. For example, Kuhs et al.[58] demonstrated experimentally that at low temperatures, stacking-disordered cubic ice gradually transforms into hexagonal ice over timescales on the order of 10 hours. Having ruled out a phase transformation, another plausible explanation for the gradual decrease in the number of grains is ongoing grain growth through coarsening and grain boundary migration. To provide quantitative evidence for this hypothesis, we analyze the evolution of the largest grain during the crystallization of liquid water containing a single INP of size L, as shown in Supplementary Figure S2(a) as a representative case. The microstructure consists of hundreds of grains, and tracking a single grain enables us to illustrate the process clearly. A general increase in the grain size is observed, consistent with coarsening. We further track all particles belonging to this grain at t =

100 ns and examine their identities at earlier times in Supplementary Figures S2(b–d). Approximately 89% of these particles were already part of the grain at t = 60 ns, while the remaining ~11% originated from other grains. These latter particles, colored blue in the figure, are located primarily along the grain boundary, providing clear quantitative evidence that the observed growth occurs through grain boundary migration.

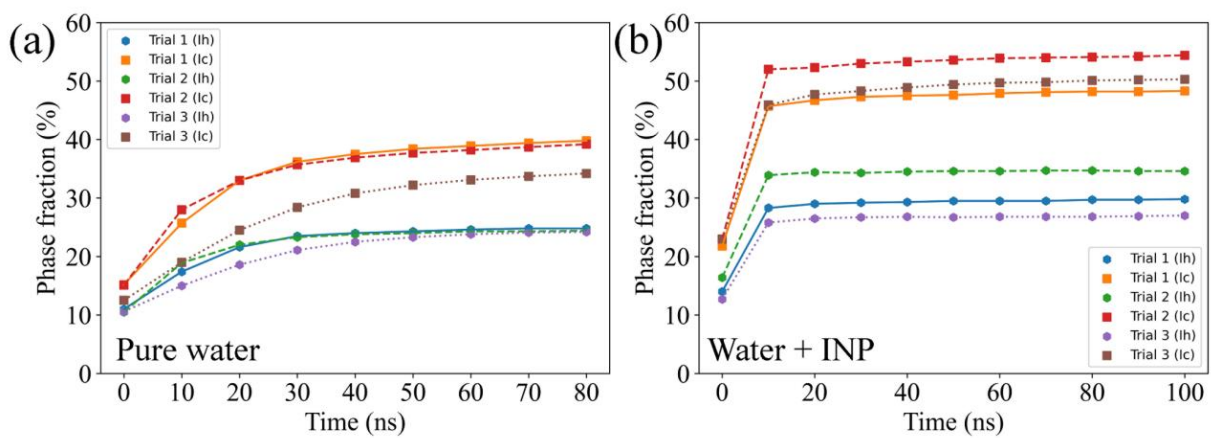

**Figure 5:** The evolution of the phase fraction of ice Ih and Ic during the crystallization of (a) liquid water, and (b) liquid water containing one INP with size of L. Time $t = 0$ represents the simulation time at which 50% of the particles in the simulation cell are characterized as 'other'-type with the ChiI+ algorithm in OVETO.

### 3.3. Effect of INPs on the creep response of ice

Here, we investigate how INPs influence the mechanical response of ice under tensile creep deformation. As shown in Figure 2, the average grain size and the number of grains in ice microstructures formed during the cooling of water vary significantly with INP density. Directly comparing the creep response of these microstructures is therefore not quantitatively meaningful, since grain structure variations alone can mask the influence of INPs and make it difficult to isolate their effect. To address this, we generate three polycrystalline ice Ih microstructures having a number of grains equal to 32, 219, and 478. These number of grains are selected to provide a representative sampling across the full range of the number of grains (32–690) for the MD simulations observed in Figure 2, enabling us to assess the effect of INP density across different number of grains. Specifically, number of grains = 32 corresponds to the minimum size, number of grains = 219 lies near the first quartile (Q1), and number of grains = 478 is close to the third quartile (Q3), ensuring coverage of low, intermediate, and high number of grains. After generating the polycrystalline ice Ih atomic structure, INPs of the desired density are introduced into the simulation cell. To prevent unphysical interactions between the INP and surrounding water molecules, all water molecules within 0.26 nm of any INP atom are removed. This threshold choice is based on the minimum distance between ice particles and INP particles within the simulation of the crystallization of liquid water containing one INP with size L, conducted in Section 3.1. It is noteworthy that our focus here is exclusively on polycrystalline ice Ih rather than a mixture of ice Ic and Ih, because ice Ih is the thermodynamically stable phase, as noted earlier.

After equilibrating the microstructures, we perform a uniaxial loading simulation to determine the yield stress, $\sigma_y$, as a reference for subsequent creep simulations. We then simulate the tensile creep at stress levels of $0.4\sigma_y$, $0.6\sigma_y$, and $0.8\sigma_y$. It is noteworthy that the temperature for all simulations is fixed at 230K, which is close to the highest ice nucleation temperature identified in the earlier section.

The steady-state creep rate $\dot{\varepsilon}$ under an applied stress $\sigma$ is often expressed as by[59]:

$$\dot{\varepsilon} = AD_0 \exp\left(-\frac{Q}{RT}\right)\left(\frac{Gb}{kT}\right)\left(\frac{b}{d}\right)^p \left(\frac{\sigma}{G}\right)^n \qquad (1)$$

where $A$ is a dimensionless constant, $D_0$ is the frequency factor, $Q$ is the activation energy of creep, $R$ is the gas constant, $T$ is the temperature, $G$ is the shear modulus, $b$ is the Burgers vector magnitude, $k$ is the Boltzmann constant, $d$ is the average grain size, and $p$ is the grain size exponent. We thus calculate the creep stress exponent $n$ from Eq. (1), as it is commonly used to infer the dominant creep deformation mechanism in materials.

The evolution of $\sigma_y$ with INP density for different numbers of grains is shown in Figure 6(a), and the corresponding stress-strain curves are provided in Supplementary Figure S3. Yield stress is determined by the 0.2% offset method, defined as the intersection of the stress–strain curve with a line parallel to the initial elastic response and offset by 0.002 strain. A general decrease in yield stress with INP density is observed at lower number of grains. This can be rationalized by the limited grain-boundary area in these microstructures, which restricts boundary-mediated plasticity. Consequently, INPs act as strong local stress concentrators, facilitating premature yielding and thereby lowering $\sigma_y$. We quantify this effect by analyzing the von Mises strain field, which highlights localized plastic deformation, in polycrystalline ice with 5 INPs (Supplementary Figure S10). The insensitivity of $\sigma_y$ to INP density for higher number of grains reflects a transition from defect-controlled yielding to grain-boundary-dominated yielding, wherein GB-mediated mechanisms weaken the incremental influence of adding more INPs. Similar behavior has been reported in nanocrystalline metals, where the grain-boundary network dominates deformation, absorbs irradiation-induced defects, and reduces the hardening contribution of additional intragranular defects, highlighting the shift in the role of grain size in controlling plastic deformation[60].

Moreover, the yield stress for nearly all INP density values decreases with increasing the number of grains. This trend is consistent with the inverse Hall–Petch relationship[61,62], where very small grain sizes (associated with a higher number of grains) can lead to reduced yield stress due to grain boundary-mediated deformation mechanisms.

It should be pointed out that the INP is considered here as a rigid body. The elastic mechanical behavior results obtained under this assumption are valid only if a small number of INP–ice bonds are broken at the yield point. A large number of broken bonds would indicate that, if the INP were instead treated as deformable, the surrounding ice atoms could experience different local stresses and structural rearrangements, suggesting that the rigid body constraint might significantly alter the mechanical response. To evaluate this, we calculate the percentage of INP–ice bonds broken at the onset of yield across all simulation conditions. The number of bonds is estimated using the 'Create Bonds' modifier in OVITO, applying pairwise cutoff distances of 1.8 times the characteristic bond length parameter from the Stillinger-Weber potential. The σ values for different atom pairs are provided in[18]. As observed in Figure 6(b), only a small fraction of bonds is broken in all cases, thereby supporting the validity of the rigid body assumption at small strains. It should be noted that the variations observed in Figure 6(b) do not reflect a systematic effect but rather arise from the random spatial distribution and orientation of the INPs in the simulations. Importantly, treating the INP under larger deformations as a perfectly rigid body may artificially constrain the surrounding ice atoms, potentially altering their structural response. This highlights that the rigid body assumption could affect the simulated ice behavior at higher strains, underscoring the importance of using interatomic potentials that accurately capture INP–ice interactions. Future work will focus on developing interatomic potentials that accurately capture INP–ice interactions.

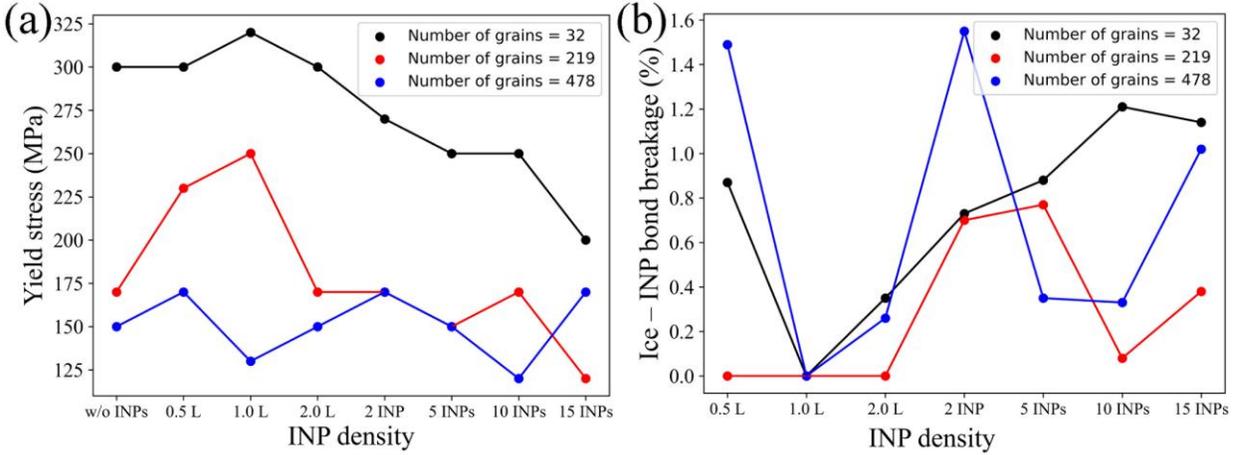

**Figure 6:** Correlation of INP density with: (a) yield stress; and (b) percentage of the number of broken bonds at the onset of yield for simulations of polycrystalline ice Ih under uniaxial loading at T = 230K.

The strain-time and corresponding strain-rate plots for the creep simulations of the polycrystalline ice Ih with number of grains = 32 and without INP are shown in Figure 7. The curves capture both primary and steady-state creep regimes, with the steady-state creep rate increasing with applied stress. To provide a complete view, the strain–time and strain-rate results for the remaining simulation settings, including different INP cases, are reported in Supplementary Figures S4–S9. Importantly, we present strain rate–strain curves for ice without INP (number of grains = 32) and ice with one INP of length L (number of grains = 32) in Supplementary Figure S12, which further confirms that steady-state creep is achieved.

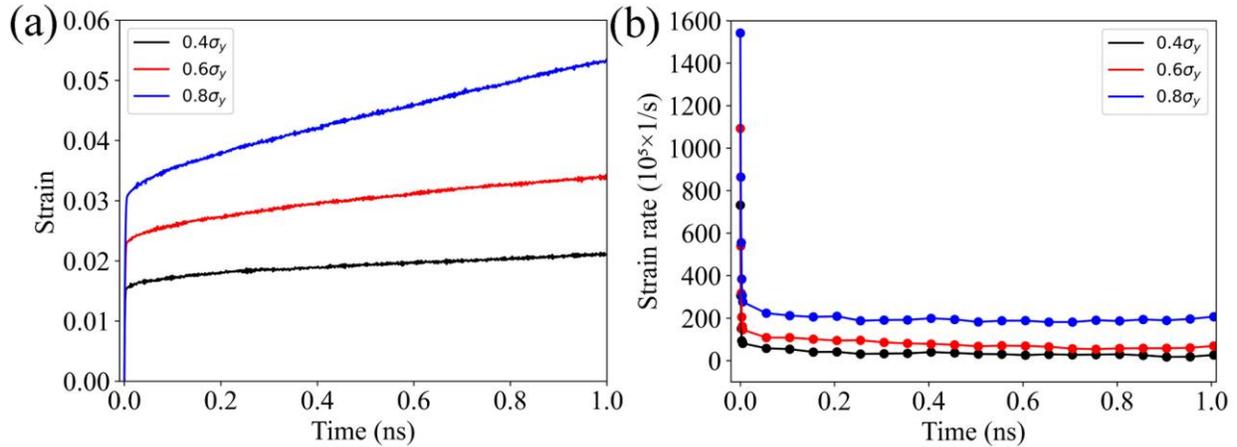

**Figure 7:** (a) Strain-time curves, and (b) Strain rate-time curve of polycrystalline ice Ih with number of grains = 32 and without INP at T=235 K under different stress loadings in a creep test.

Using the strain-rate data shown in Figure 7(b) for pure ice (without INP), along with corresponding data for simulations with varying INP densities reported in Supplementary Figure S4-S9, we then calculate the stress exponent $n$ in Equation (1). The results are summarized in Figure 8. It is observed that the stress exponent $n$ falls within the range of $1 < n < 3$ for nearly all INP densities and number of grains. This range suggests that the dominant creep mechanism in the ice across all number of grains and INP densities investigated in this study is diffusion-accommodated grain boundary sliding rather than dislocation activity. The only exception is polycrystalline ice Ih without INPs and with number of grains = 32, where the stress exponent is calculated as $n = 3.02$. This case will be examined in more detail later in this section, where we analyze the corresponding dislocation evolution to clarify the underlying mechanism.

The observation that the stress exponent $n$ falls between 1 and 3 across nearly all INP densities and grain numbers suggests that grain boundary sliding governs the creep behavior. This interpretation is supported by von Mises shear strain calculation for a representative simulation of polycrystalline ice with 1 INP and with number of grains = 32 (Supplementary Figure S11), which shows that the plasticity localizes along grain boundaries. Furthermore, the ratio of mean-squared-displacement of grain boundary atoms over the bulk atoms (R), calculated over two subsequent time frames at 0.5 ns and 0.55 ns in the steady-state portion of the creep, reaches $R = 1.25$, indicating that grain-boundary atoms are more mobile than bulk atoms. This enhanced mobility allows the boundaries to accommodate the relative sliding of grains, providing further evidence that the observed deformation is diffusion-accommodated grain boundary sliding. Importantly, stress exponent *n* decreases with increasing the number of grains for nearly all simulations which can be attributed to the growing fraction of grain-boundary atoms in finer-grained samples that increases the contribution of this mechanism to overall deformation. A similar decrease in the stress exponent with decreasing grain size has been observed in slow snow compression[63].

Another observation from the figure is that the stress exponent shows an almost decreasing trend with increasing INP density for the number of grains = 32. This trend likely arises because the INPs generate local stress concentrations near grain boundaries, promoting diffusion-accommodated grain-boundary sliding. This effect is not observed in finer-grained samples, where grain-boundary sliding already dominates the creep response.

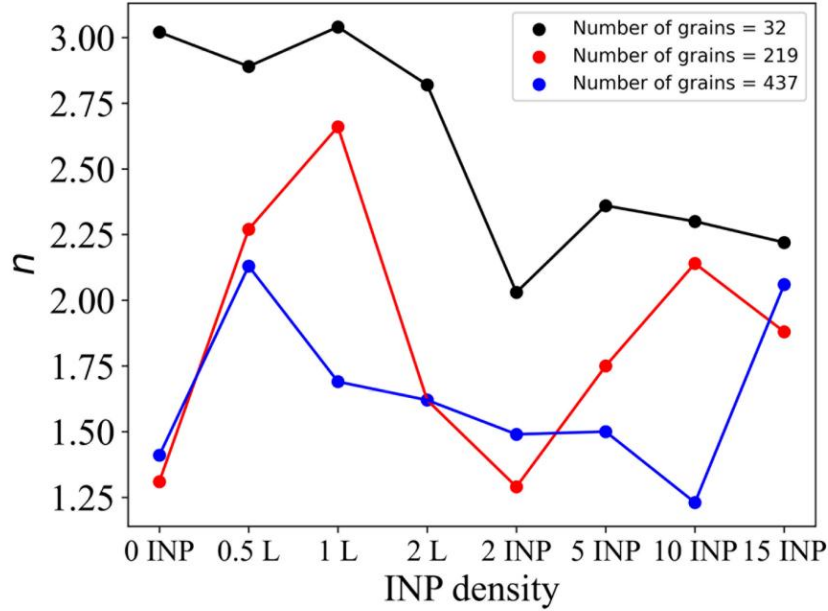

**Figure 8:** Correlation of density of INPs with the exponent parameter *n* for different number of grains.

As mentioned earlier, for polycrystalline ice without INPs and with a number of grains = 32, the stress exponent *n* = 3.02, indicating dislocation-driven creep. To better understand this behavior, we compare the dislocation networks of this case with those of polycrystalline ice containing 15 INPs, also with a number of grains = 32. Experimentally, the dominant dislocations in ice are the a-type $<11\bar{2}0>$ and c-type $<0001>$ dislocations[64]. As shown in Figure 9(a-b), creep loading in the INP-free case produces substantial dislocation evolution: the number of $<11\bar{2}0>$ segments decreases from 44 to 9, and $<0001>$ segments from 2 to 0. In contrast, the dislocation network in the presence of 15 INPs, shown in Figure 9(c-d), exhibits only minor changes, with $<11\bar{2}0>$ segments decreasing from 36 to 33 and $<0001>$ segments from 3 to 0. This apparent difference suggests that, without INPs, plastic deformation is primarily accommodated by dislocation activity, consistent with the higher stress exponent (n = 3.02). In contrast, the presence of INPs suppresses dislocation motion and promotes diffusion-accommodated grain boundary sliding, characterized by a lower stress exponent (*n* = 2.22).

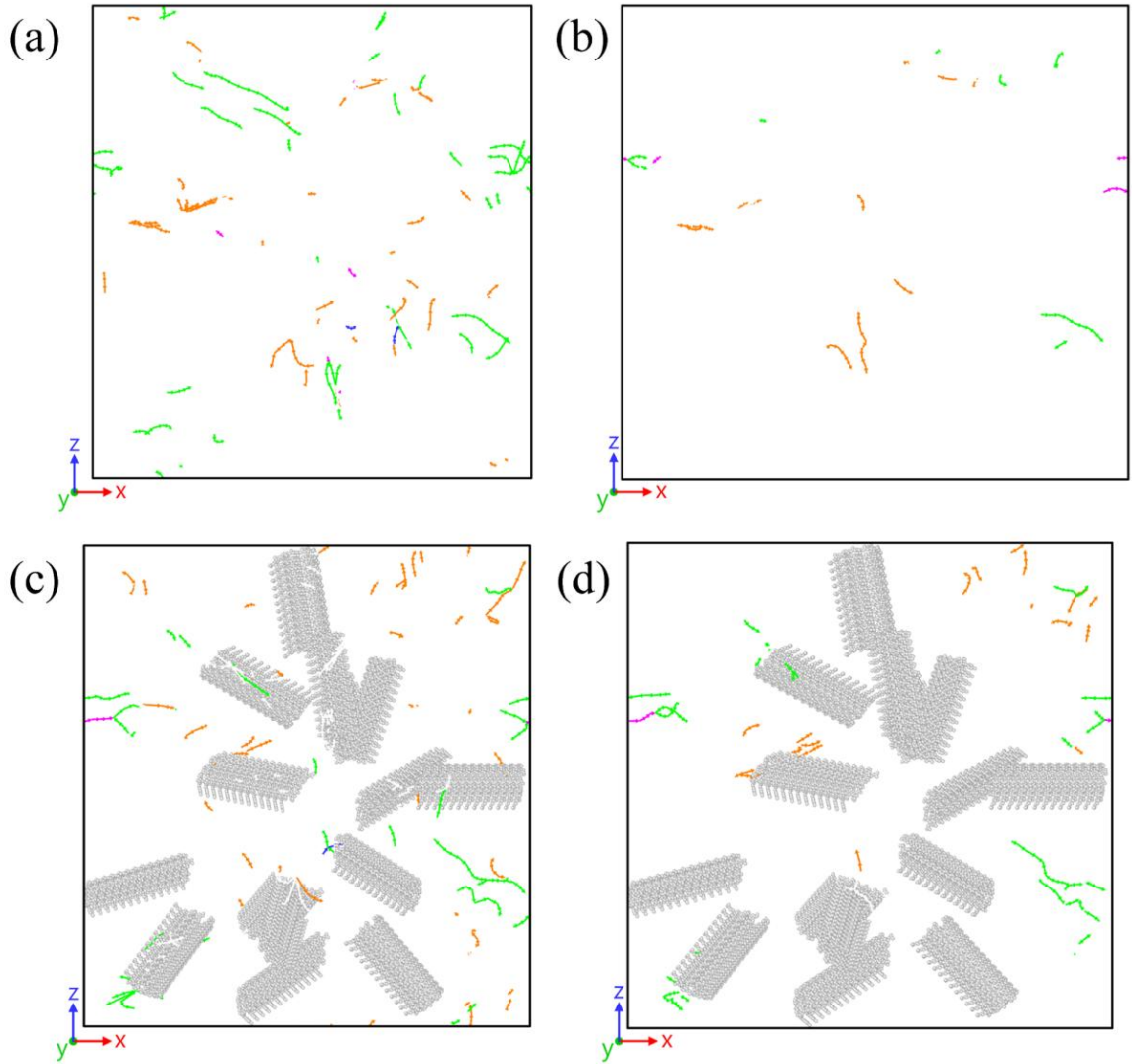

**Figure 9**: Microstructures of (a–b) polycrystalline ice without INPs and (c–d) polycrystalline ice with 15 INPs, where (a) and (c) correspond to *t* = 0 and (b) and (d) to *t* = 2.5 ns after applying the creep load. Colored thin tubes represent dislocation lines: $\frac{1}{3} < 1\bar{2}10 >$ in green, $< 0001 >$ in blue, $< 1\bar{1}00 >$ in pink and $\frac{1}{3} < 1\bar{1}00 >$ in orange. For clarity, dislocations labeled as 'Other' are not shown.

# 4. Conclusion

In conclusion, we used molecular dynamics (MD) simulations to investigate how varying INP densities influenced the ice nucleation temperature, the resulting ice microstructure, and the mechanical behavior of the ice under creep tensile loading. The results of MD simulations revealed that the inclusion of INPs could raise the ice nucleation temperature of liquid water above the homogeneous value by up to 23K. Our results further showed that the ice nucleation temperature initially increased with INP density and then plateaued due to the limited lateral stabilization area of INP binding sites. At higher densities, the temperature decreased as overlapping ice-promoting regions disrupted the surrounding water structure, reducing the available free volume for nucleation.

Upon comparing the microstructures of ice formed at different initial INP densities, we found that the number of grains initially decreased with increasing INP density, promoting crystal growth from fewer, larger nucleation sites. At higher densities, however, the number of grains increased due to the simultaneous formation of multiple independent grains. Heterogeneous nucleation also led to faster stabilization of the grain structure as compared to homogeneous nucleation, due to reduced nucleation barriers at higher temperatures. While the number of grains continued to decrease slightly after heterogeneous nucleation, no significant Ic-to-Ih phase transformation was observed that could account for this behavior. Instead, this trend was attributed to grain coarsening driven by grain-boundary migration.

Another notable observation was that INP density exhibited an inverse relationship with yield stress, particularly at a lower number of grains, due to the structural defects and local stress concentrations introduced by the INPs. Yield stress also decreased with increasing the number of grains, consistent with the inverse Hall–Petch effect. Importantly, analysis of the stress exponent revealed that diffusion-accommodated grain boundary sliding dominated across INP densities ($1 < n < 3$). In the absence of INPs and especially at lower number of grains, dislocation activity governed deformation, as indicated by a higher stress exponent ($n = 3.02$) and noticeable dislocation network evolution. By contrast, INP-containing systems showed limited dislocation activity and lower stress exponents, confirming a transition to diffusion-accommodated grain boundary sliding creep mechanisms. Overall, our study highlights that INPs can be used to tune both ice nucleation dynamics and mechanical behavior. By controlling the nucleation rate, resulting microstructure, and mechanical response, INPs offers a strategy for designing ice with tailored durability and resilience in real-world settings, including polar environments, engineered ice systems, and other environments where ice stability is critical. These findings underscore the broader impact of understanding and manipulating ice mechanics through molecular-level interventions.

# Acknowledgment

This work was sponsored by the Seaver Institute through grant number TSI-2025-M. The authors acknowledge grant support from the Hopkins Extreme Materials Institute (HEMI) and the Research and Exploratory Development (RED) mission area of the Johns Hopkins Applied Physics Laboratory (JHU/APL). C. D. S. gratefully acknowledges internal financial support from the Johns Hopkins Applied Physics Laboratory's Independent Research & Development (IR&D) Program. Computational resources were provided by the Advanced Research Computing at Hopkins (ARCH).

# Declaration of Competing Interest

The authors declare that they have no known competing financial interests or personal relationships that could have appeared to influence the study reported in this paper.

# Data Availability

All data generated, used and/or analyzed during the current study are available from the corresponding author on request.

# Supplementary Material for

# Effect of ice nucleating proteins on the structure-property relationships of ice: A molecular dynamics study

Ali K. Shargh[1*], Christopher D. Stiles[1,2], Jaafar A. El-Awady[1*]

[1] Department of Mechanical Engineering, Johns Hopkins University, Baltimore, Maryland 21218, U.S.A.

[2] Research and Exploratory Development Department, Johns Hopkins Applied Physics Laboratory, Laurel, Maryland 20723, U.S.A.

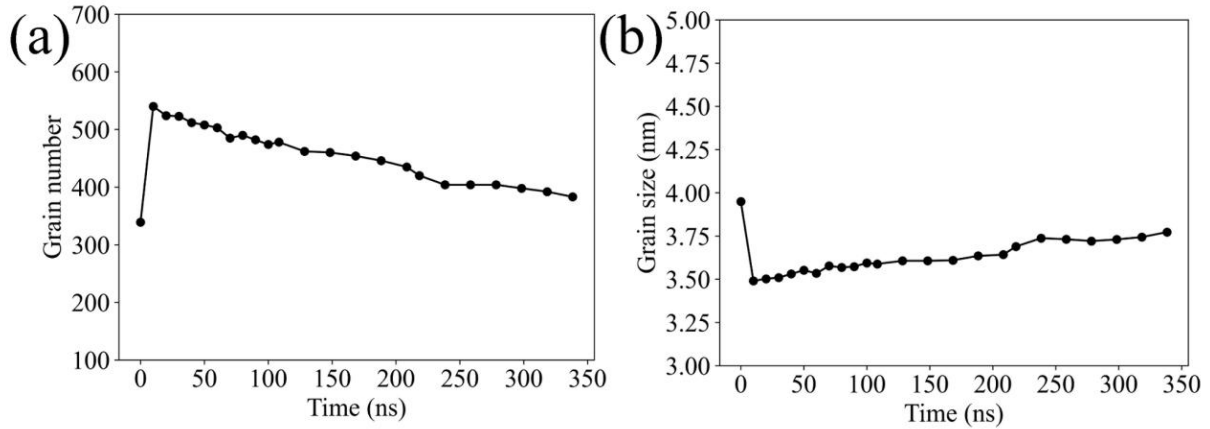

**Figure S1**: Evolution of (a) number of grains, and (b) grain size during the crystallization of liquid water containing one INP with size of $L$ in the final equilibration steps after the formation of a stable ice cluster under cooling with an excessively long equilibration step. The simulation is kept

---

[*] Corresponding authors:
  Email addresses: ashargh1@jhu.edu (A. K. Shargh), jelawady@jhu.edu (J. A. El-Awady)

at temperatures of 226 K. Note that in the first frame, 50% of particles are characterized as 'other' with Chil+ algorithm. The total time of all simulations is 130ns.

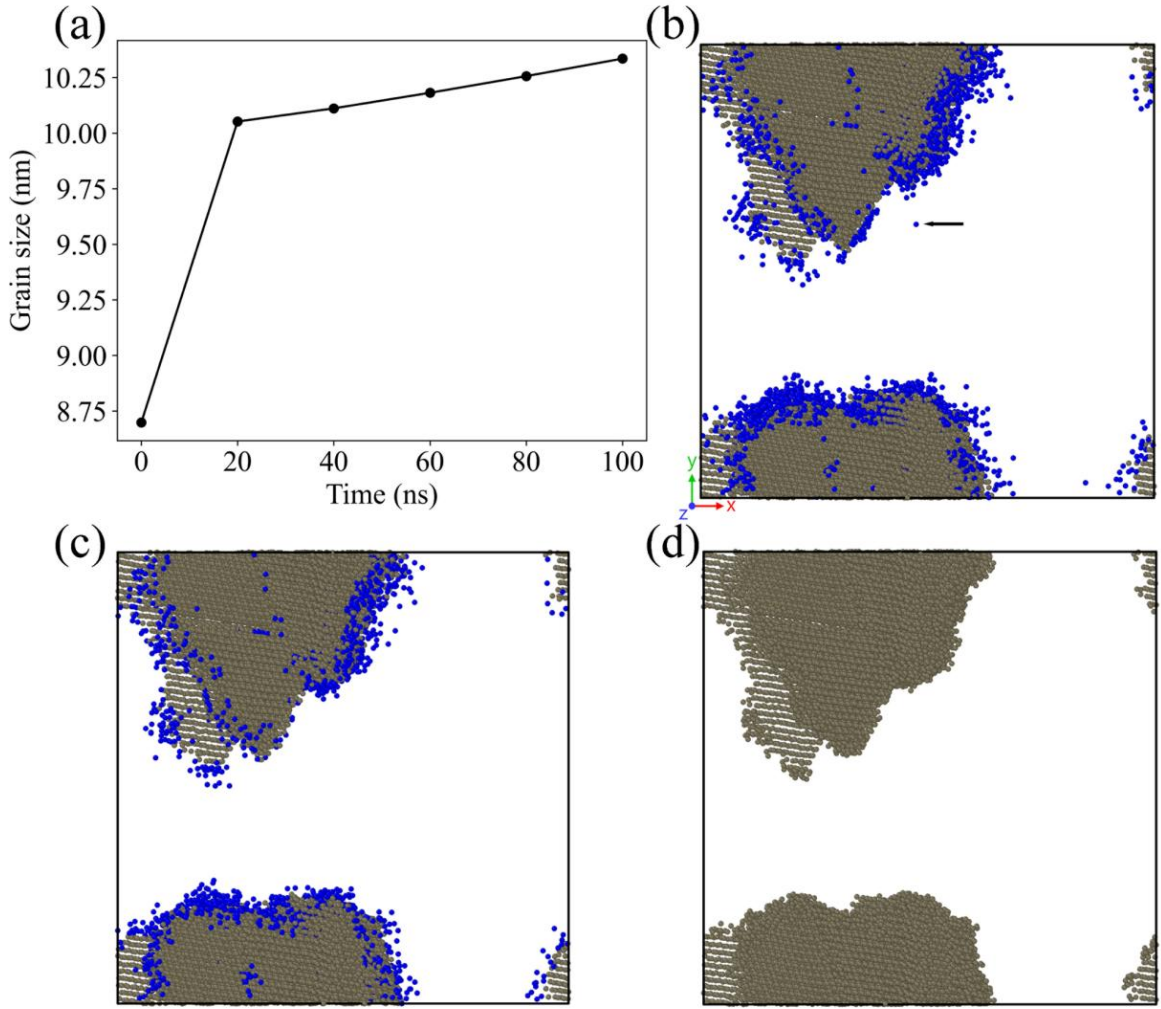

**Figure S2**: (a) Evolution of the grain size of the largest grain during the crystallization of liquid water containing a single INP of size $L$. (b-d) Snapshots of this largest grain at selected times: (b) $t = 60$ ns, (c) $t = 80$ ns, and (d) $t = 100$ ns. Particles colored brown belong to grain ID = 1 (i.e., the largest grain) at t = 100 ns, while particles colored blue belongs to grain ID = 1 at $t = 100$ ns but were part of other grains at earlier times. This figure is representative and highlights continuous grain growth within the ice microstructure via coarsening and grain boundary migration. Note that particles far from the grain boundary, such as the one highlighted with the arrow in (b), are in a non-crystalline state.

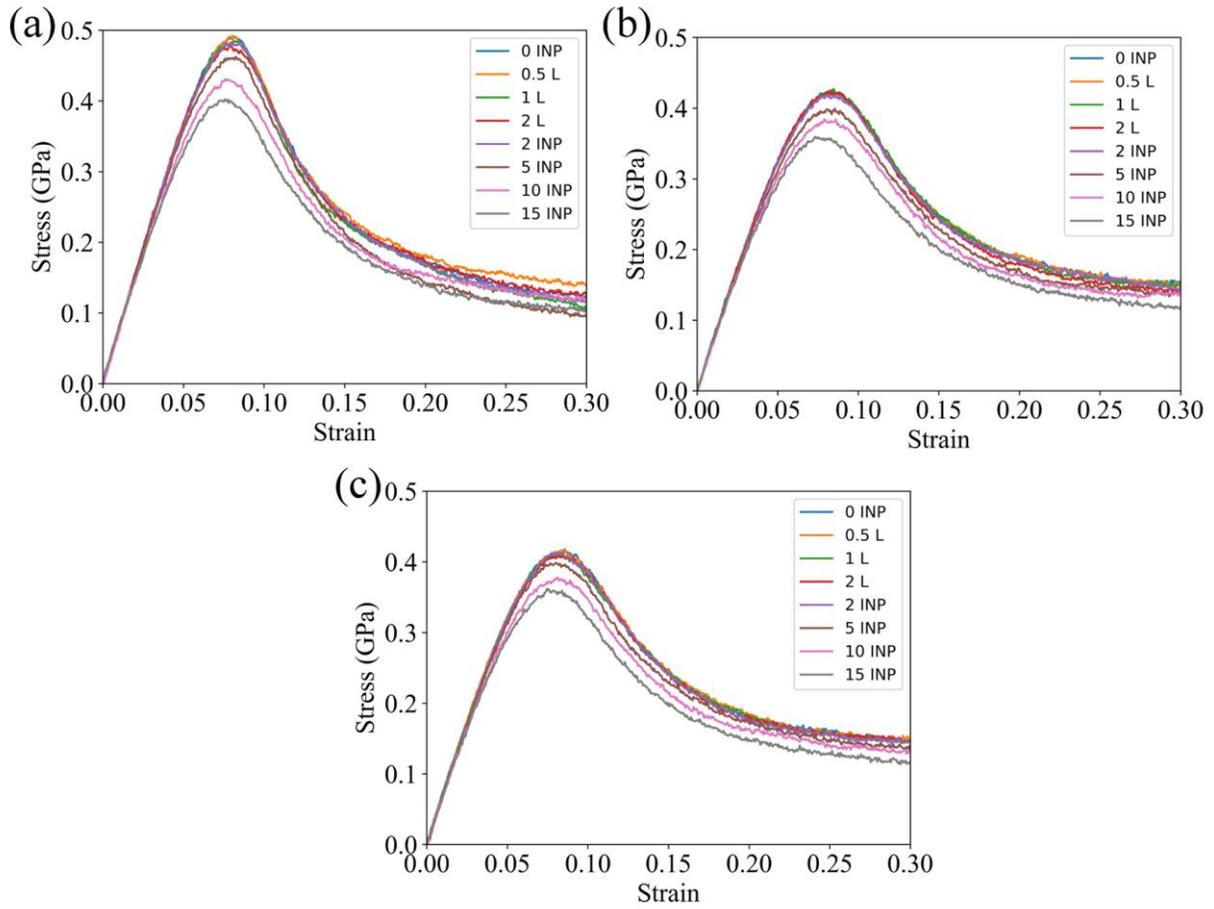

**Figure S3**: Stress-strain curves of polycrystalline ice Ih with varying density of INPs under uniaxial loading for (a) number of grains = 32, (b) number of grains = 219, and (c) number of grains = 478

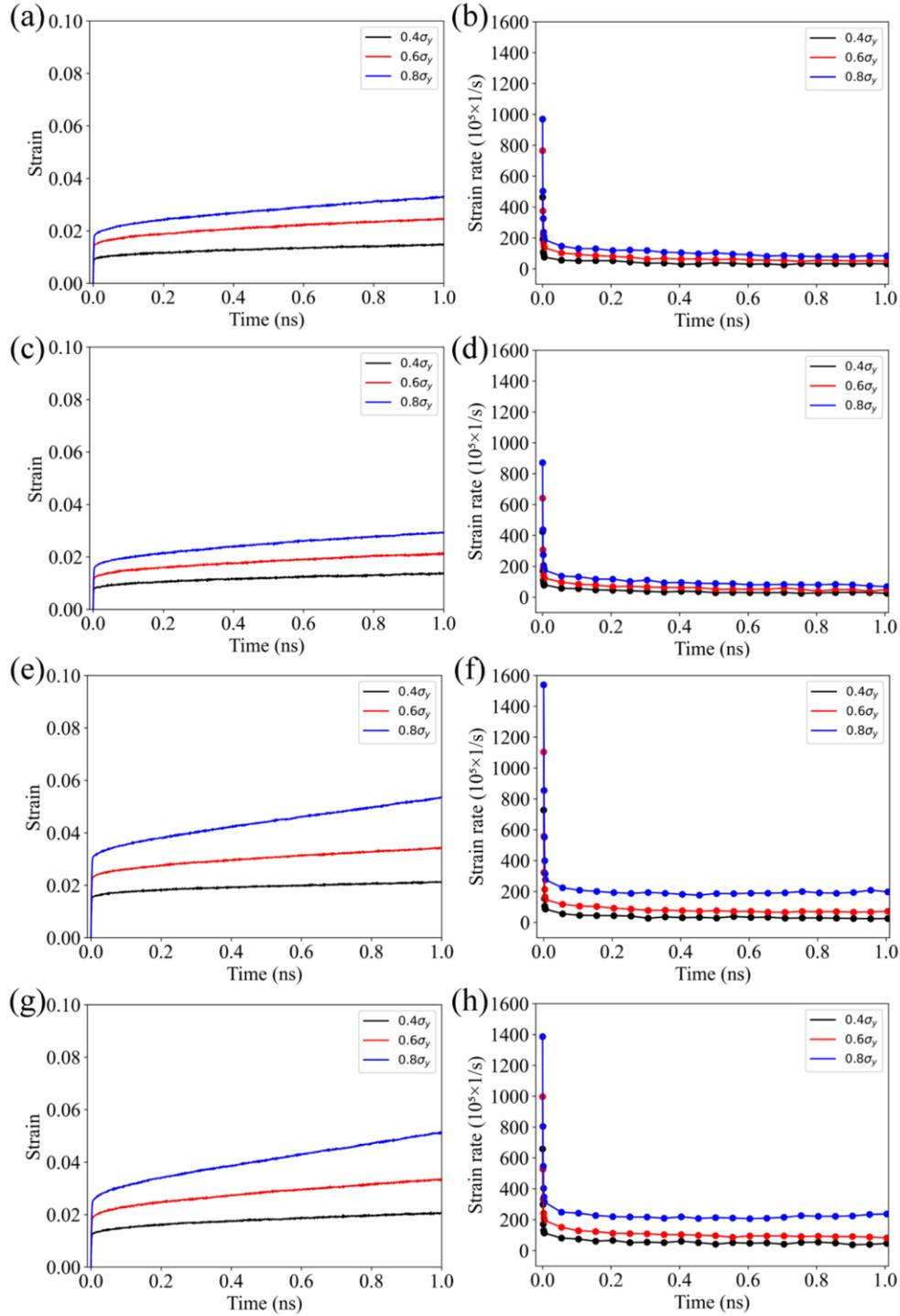

**Figure S4**: Strain-time curves and corresponding strain rate-time curve of (a-b) ice with number of grains = 219 and without INP, (c-d) ice with number of grains = 479 and without INP, (e-f) ice with number of grains = 32 and with 1 INP with size 0.5*L*, (g-h) ice with number of grains = 219 and with 1 INP with size 0.5*L* at *T* = 235 K under different stress loadings in a creep test.

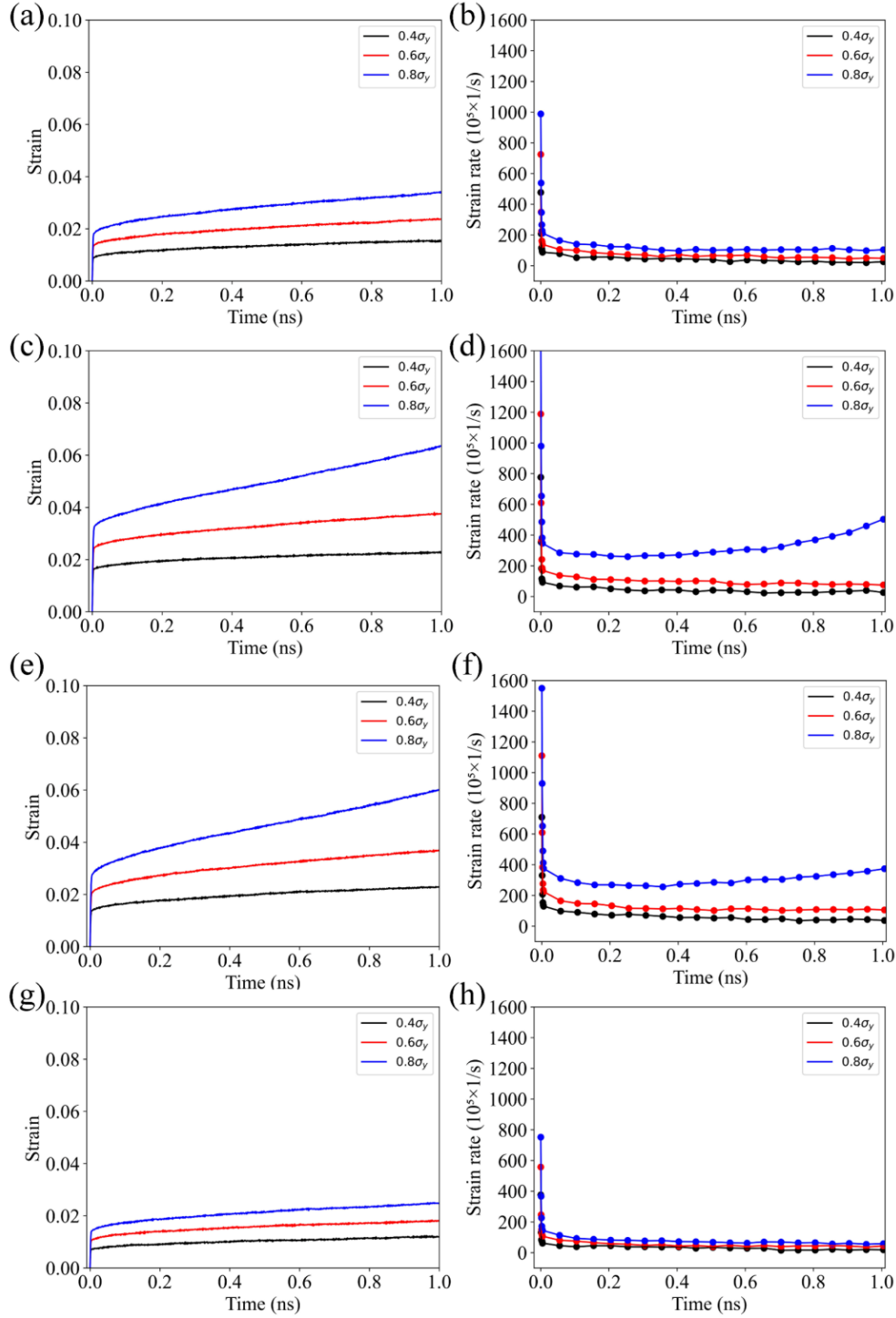

**Figure S5**: Strain-time curves and corresponding strain rate-time curve of (a-b) ice with number of grains = 478 and with 1 INP with size 0.5$L$, (c-d) ice with number of grains = 32 and with 1 INP with size $L$, (e-f) ice with number of grains = 219 and with 1 INP with size $L$, (g-h) ice with number of grains = 478 and with 1 INP with size $L$ at $T$ = 235 K under different stress loadings in a creep test.

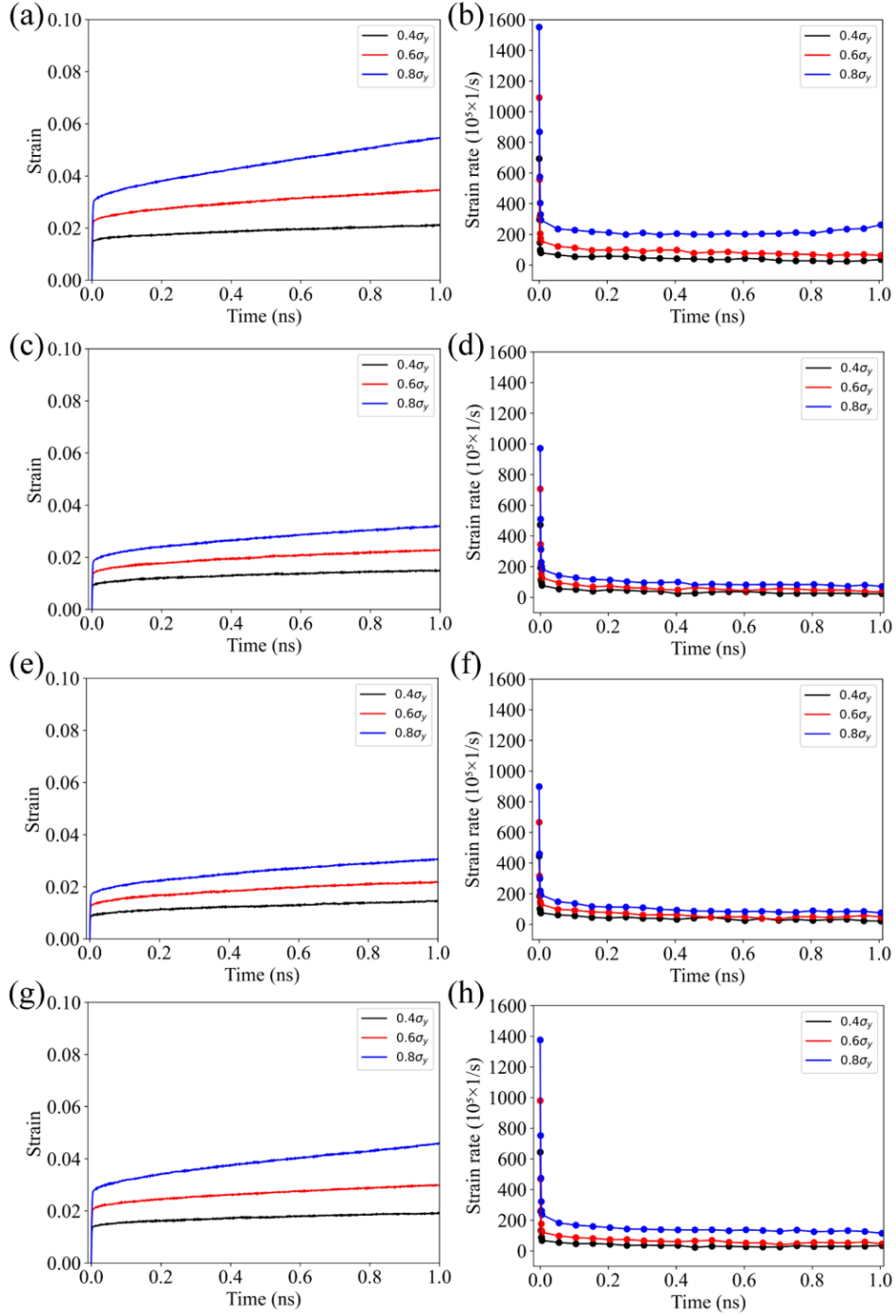

**Figure S6**: Strain-time curves and corresponding strain rate-time curve of (a-b) ice with number of grains = 32 and with 1 INP with size 2$L$, (c-d) ice with number of grains = 219 and with 1 INP with size 2$L$, (e-f) ice with number of grains = 478 and with 1 INP with size 2$L$, (g-h) ice with number of grains = 32 and with 2 INP at $T$ = 235 K under different stress loadings in a creep test.

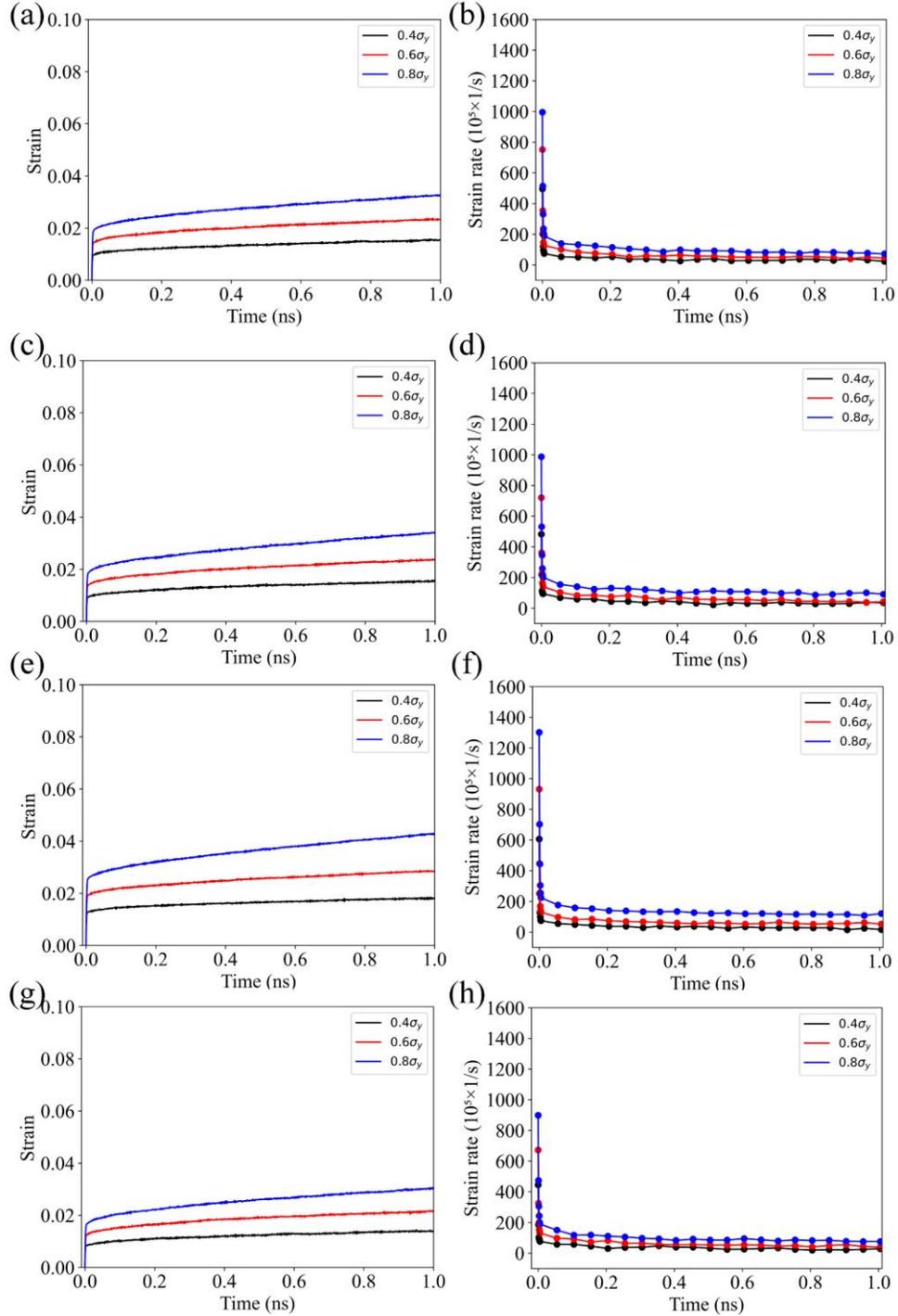

**Figure S7**: Strain-time curves and corresponding strain rate-time curve of (a-b) ice with number of grains = 219 and with 2 INP, (c-d) ice with number of grains = 478 and with 2 INP, (e-f) ice with number of grains = 32 and with 5 INP, (g-h) ice with number of grains = 219 and with 5 INP at $T$ = 235 K under different stress loadings in a creep test.

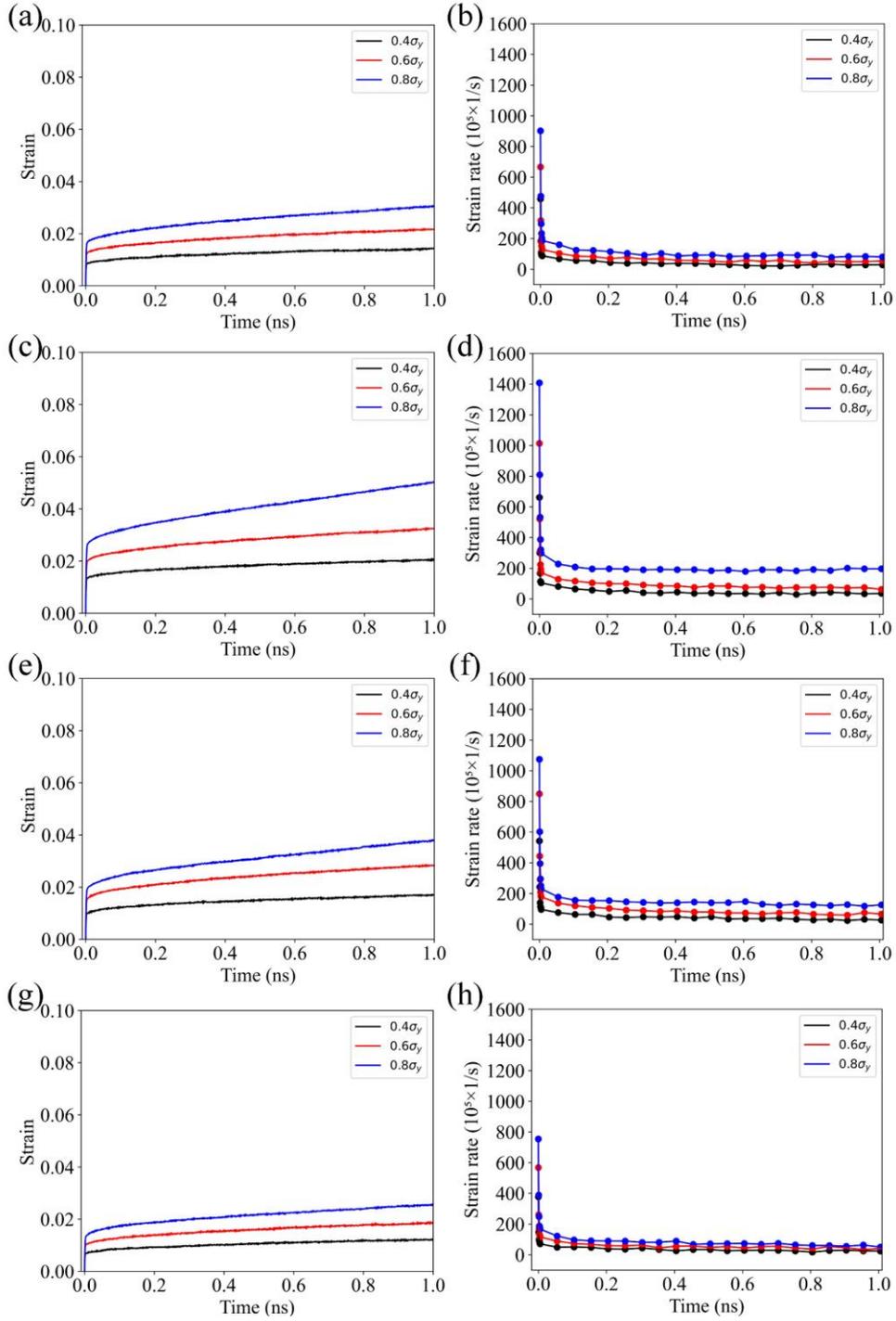

**Figure S8**: Strain-time curves and corresponding strain rate-time curve of (a-b) ice with number of grains = 478 and with 5 INP, (c-d) ice with number of grains = 32 and with 10 INP, (e-f) ice with number of grains = 219 and with 10 INP, (g-h) ice with number of grains = 478 and with 10 INP at $T$ = 235 K under different stress loadings in a creep test.

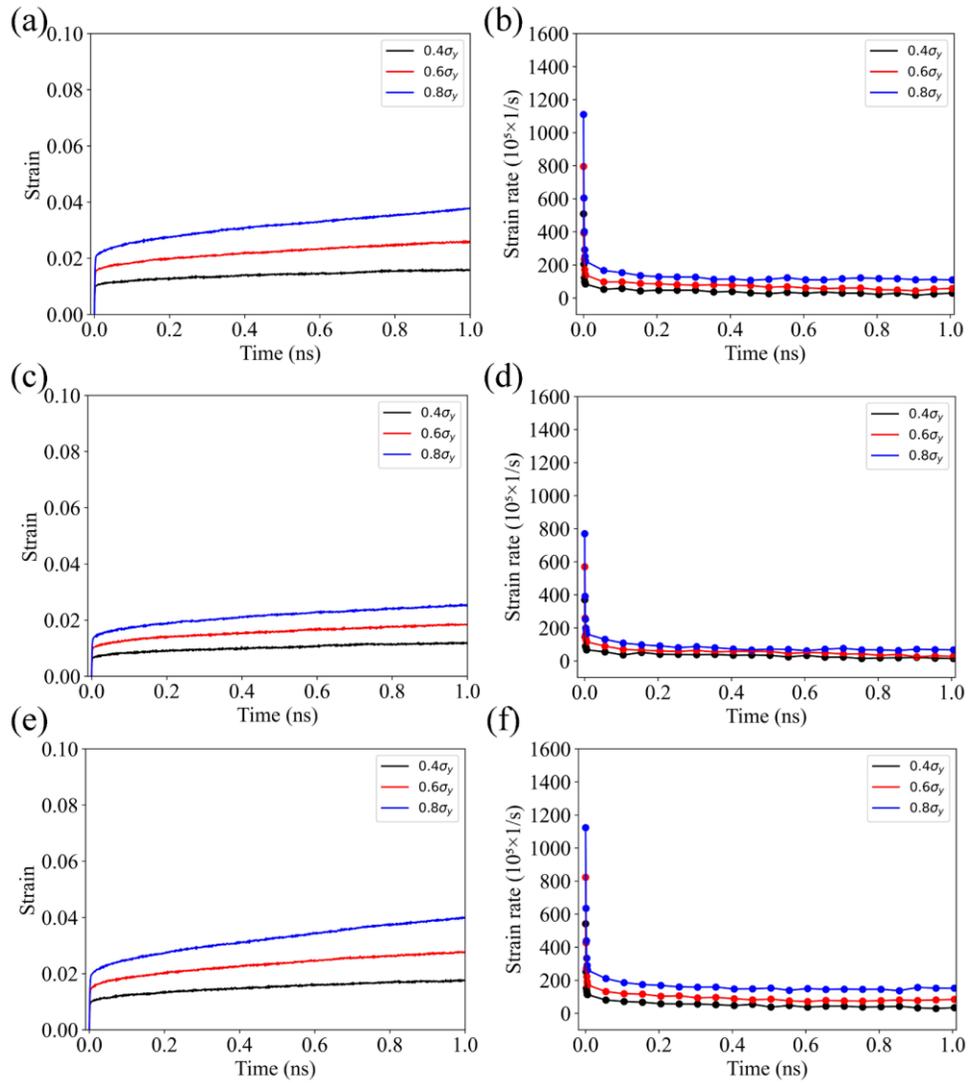

**Figure S9**: Strain-time curves and corresponding strain rate-time curve of (a-b) ice with number of grains = 32 and with 15 INP, (c-d) ice with number of grains = 219 and with 15 INP, (e-f) ice with number of grains = 478 and with 15 INP at $T$ = 235 K under different stress loadings in a creep test.

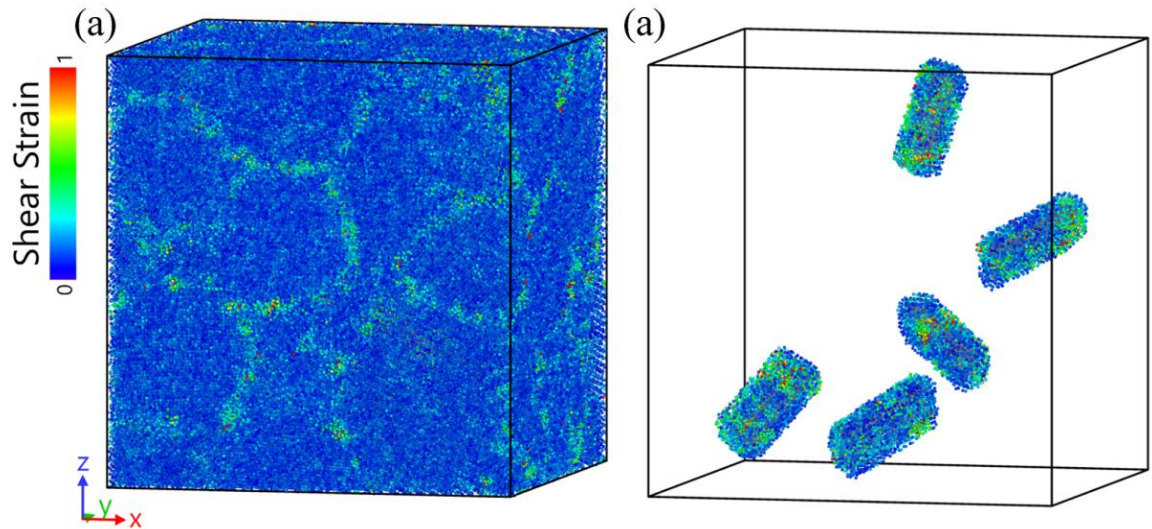

**Figure S10**: (a-b) Visualization of the Von Mises shear strain distribution in polycrystalline ice containing 5 INPs, subjected to uniaxial tensile loading along the *x*-axis at a strain = 0.025. In (b), ice particles surrounding the INPs are highlighted by selecting all particles within a cutoff distance of 0.8 nm from each INP.

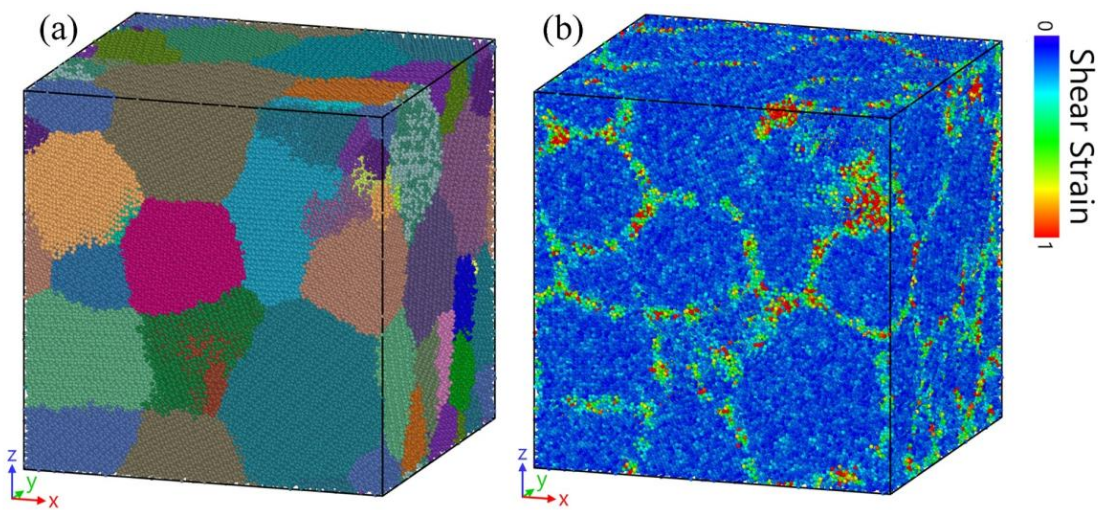

**Figure S11**: Visualization of (a) Grain distribution and (b) the Von Mises shear strain distribution in polycrystalline ice containing 1 INP with size *L*, subjected to creep test along the *x*-axis at time = 0.5 ns.

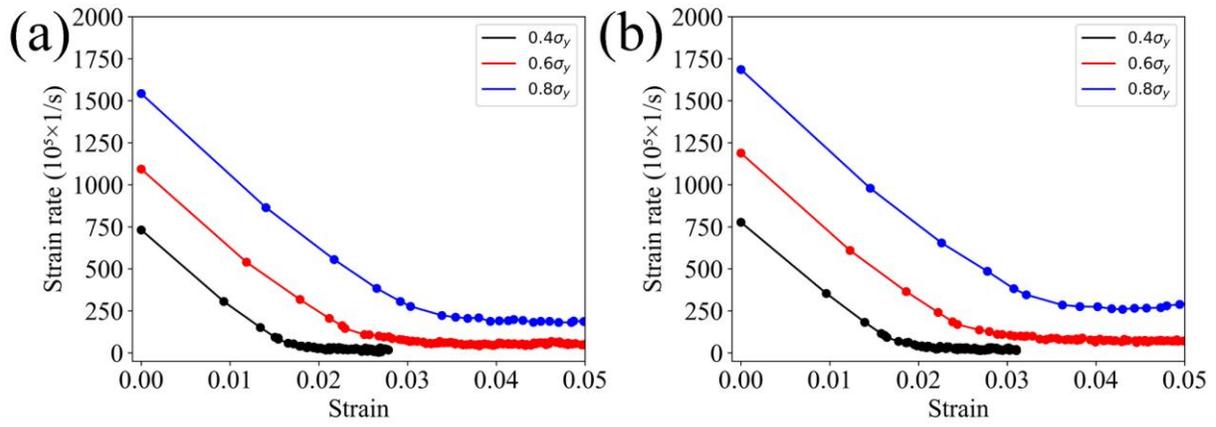

**Figure S12**: Strain rate-strain curves of (a) ice with number of grains = 32 and without INP, (b) ice with number of grains = 32 and with 1 INP of length L at $T$ = 235 K under different stress loadings in a creep test.